\documentclass[sigconf]{acmart}

\AtBeginDocument{%
  \providecommand\BibTeX{{%
    \normalfont B\kern-0.5em{\scshape i\kern-0.25em b}\kern-0.8em\TeX}}}

\setcopyright{acmcopyright}
\copyrightyear{2018}
\acmYear{2018}
\acmDOI{10.1145/1122445.1122456}

\setcopyright{none}

\acmConference[Woodstock '18]{Woodstock '18: ACM Symposium on Neural
  Gaze Detection}{June 03--05, 2018}{Woodstock, NY}
\acmBooktitle{Woodstock '18: ACM Symposium on Neural Gaze Detection,
  June 03--05, 2018, Woodstock, NY}
\acmPrice{15.00}
\acmISBN{978-1-4503-XXXX-X/18/06}

\usepackage{multirow}
\usepackage{footmisc}
\usepackage{subcaption,siunitx,booktabs}
\usepackage{caption}
\usepackage{bbm}
\usepackage{tablefootnote}

\usepackage{multirow}
\usepackage{balance}
\usepackage{enumitem}

\begin{document}

\title{CCL4Rec: Contrast over Contrastive Learning for Micro-video Recommendation}

\author[S. Zhang*, B. Li*, D. Yao*, F. Feng, J. Zhu, W. Fan, Z. Zhao, X. He, T. Chua,  F. Wu]{
    Shengyu Zhang$^{1*}$, Bofang Li$^{1*}$, Dong Yao$^{1*}$, Fuli Feng$^{4}$, Jieming Zhu$^{6}$, Wenyan Fan$^{1}$, Zhou Zhao$^{1,2}$, \\ Xiaofei He$^{1}$, Tat-seng Chua$^{5}$, Fei Wu$^{1,2,3}$
}
\affiliation{
    $^1$ Zhejiang University \ $^2$ Shanghai Institute for Advanced Study of Zhejiang University \country{}
}
\affiliation{
	$^3$ Shanghai AI Laboratory \ $^4$ University of Science and Technology of China\country{}
}
\affiliation{
	$^5$ National University of Singapore \ $^6$ Huawei Noah's Ark Lab\country{}
}

\email{
  {sy_zhang, yaodongai, zhaozhou, wufei}@zju.edu.cn
}

\email{
fulifeng93@gmail.com, jamie.zhu@huawei.com, chuats@comp.nus.edu.sg
}

\email{
	wenyan.17@intl.zju.edu.cn, chuxuepsn@163.com, xiaofei_h@qq.com
}


\renewcommand{\shortauthors}{Shengyu Zhang, et al.}
\newcommand{\etal}{\textit{et al}.}
\newcommand{\ie}{\textit{i}.\textit{e}.}
\newcommand{\eg}{\textit{e}.\textit{g}.}
\newcommand{\wrt}{\textit{w}.\textit{r}.\textit{t}. }
\newcommand{\vpara}[1]{\vspace{0.05in}\noindent\textbf{#1 }}



\begin{abstract}

Micro-video recommender systems suffer from the ubiquitous noises in users' behaviors, which might render the learned user representation indiscriminating, and lead to trivial recommendations (\eg, popular items) or even weird ones that are far beyond users' interests. Contrastive learning is an emergent technique for learning discriminating representations with random data augmentations. However, due to neglecting the noises in user behaviors and treating all augmented samples equally, the existing contrastive learning framework is insufficient for learning discriminating user representations in recommendation. To bridge this research gap, we propose the \textit{Contrast over Contrastive Learning} framework for training recommender models, named CCL4Rec, which models the nuances of different augmented views by further contrasting augmented positives/negatives with adaptive pulling/pushing strengths, \ie, the contrast over (vanilla) contrastive learning. To accommodate these contrasts, we devise the hardness-aware augmentations that track the importance of behaviors being replaced in the query user and the relatedness of substitutes, and thus determining the quality of augmented positives/negatives. The hardness-aware augmentation also permits controllable contrastive learning, leading to performance gains and robust training. In this way, CCL4Rec captures the nuances of historical behaviors for a given user, which explicitly shields off the learned user representation from the effects of noisy behaviors. We conduct extensive experiments on two micro-video recommendation benchmarks, which demonstrate that CCL4Rec with far less model parameters could achieve comparable performance to existing state-of-the-art method, and improve the training/inference speed by several orders of magnitude.


\end{abstract}




\maketitle

\section{Introduction}

Micro-video online services, such as TikTok, Kuaishou, and Instagram, have developed rapidly in recent years, leading to the proliferation of micro-video production and communication. 
The tremendously grown volume of micro-videos has intensified the need of retrieval and recommender systems that can permit personalized content discovery and consumption in micro-video online services. In the literature of modern recommender systems, the recent emergence of deep learning techniques has provided enticing methods \cite{Hidayati_Hsu_Chang_Hua_Fu_Cheng_2018,Jiang_Wang_Liu_Nie_Duan_Xu_2019,Wang_Feng_He_Nie_Chua_2020,Wei_Cheng_Yu_Zhao_Zhu_Nie_2019,Wei_Wang_Nie_He_Hong_Chua_2019,Gao_Zhang_Xu_2017,Li_Liu_Yin_Cui_Xu_Nie_2019,Chen_Liu_Zha_Zhou_Xiong_Li_2018,Jiang_Wang_Wei_Gao_Wang_Nie_2020,Du_Wang_He_Li_Tang_Chua_2020,Li_Xu_Jiang_Cao_Huang_2020,Wei_Wang_Nie_He_Chua_2020,Yu_Gan_Wei_Cheng_Nie_2020,Yang_Xie_Wang_Yuan_Ding_Yan_2020,Lu_Zhang_Huang_Wang_Yu_Zhao_Wu_2020,DBLP:conf/kdd/QianXLZJLZC022} in learning user representations that capture the characteristics of users' preferences and interests.
 More recently, in frameworks that are specifically designed for micro-video recommendation, we notice that Li \etal, \cite{Li_Liu_Yin_Cui_Xu_Nie_2019} encapsulate the temporally remote connections upon the behavior sequence based on visual similarity and devise the temporal graph-based LSTM as the preference encoder. Jiang \etal, \cite{Jiang_Wang_Wei_Gao_Wang_Nie_2020} explicitly perform group routing and assignment to capture group-level interest. Many of these works follow the sequential recommendation schema \cite{Huang_Qian_Fang_Sang_Xu_2018,Huang_Fang_Qian_Sang_Li_Xu_2019,Rendle_Freudenthaler_SchmidtThieme_2010,Hidasi_Karatzoglou_Baltrunas_Tikk_2016,Zhou_Zhu_Song_Fan_Zhu_Ma_Yan_Jin_Li_Gai_2018}, which typically makes predictions based on the historical behavior sequence and a sequence modeling module.


\begin{figure}[!t] \begin{center}
    \includegraphics[width=\columnwidth]{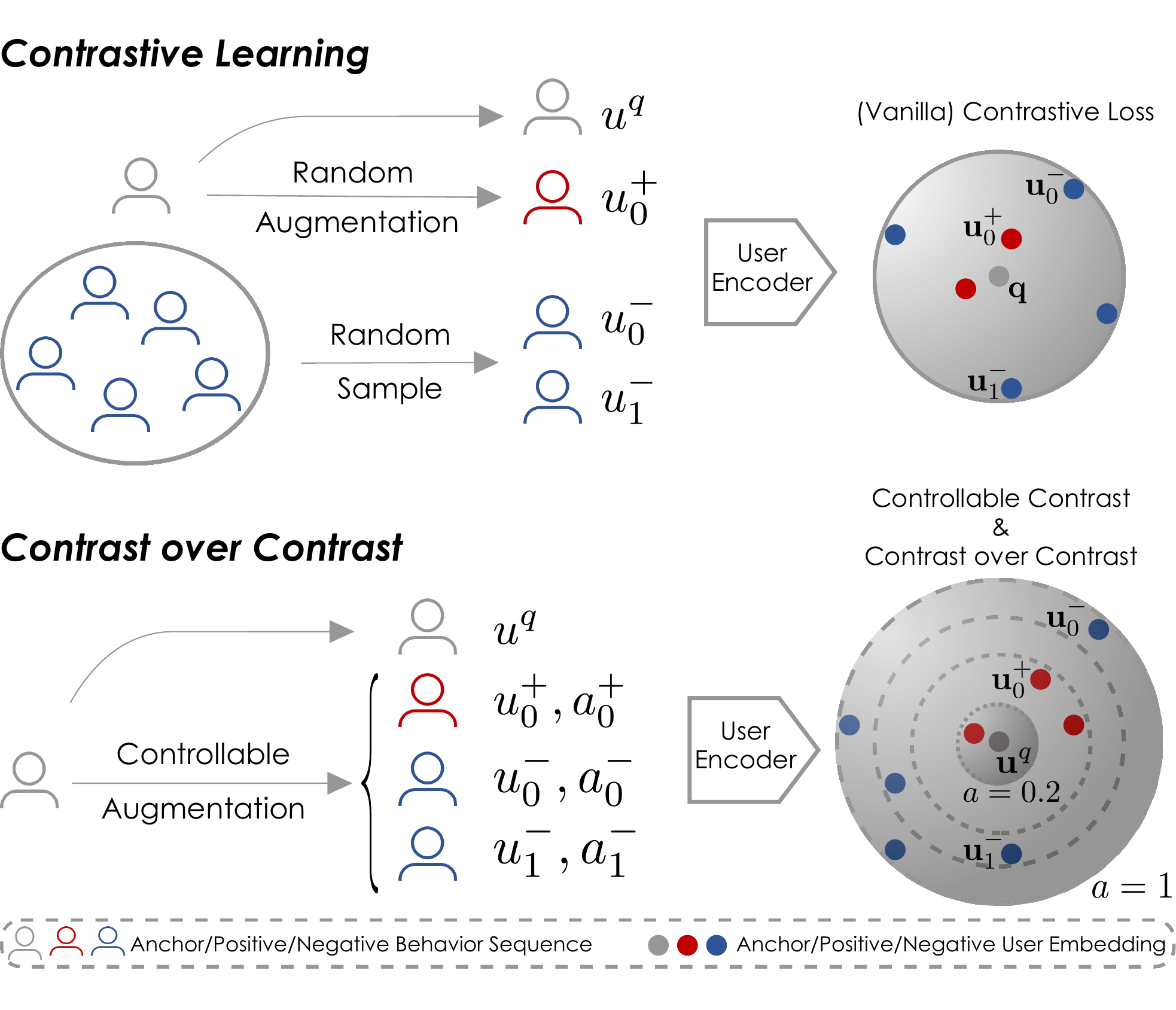}
    \caption{
		An illustration of the proposed contrast over contrast framework, which explicitly models the hardness/quality $a$ of augmented samples and the nuances between them to learn discriminating user representations.
    	}
\label{fig:firstpage}
\end{center} \end{figure}

\begin{sloppypar}
	Despite the fruitful progress in the literature, it is worth noting that there are still some common challenges for micro-video recommendation. One of the major challenges can be the sheer difficulty of learning discriminating and effective user representations. Specifically, users behaviors are mostly implicit feedbacks and the ubiquitous false-positive interactions \cite{Wang_Feng_He_Nie_Chua_2020} may render the learned user representation indiscriminating (\ie, towards an average/over-smoothing representation of all users). Recently, contrastive learning \cite{DBLP:conf/sigir/ZhangYZC021,DBLP:conf/www/ZhangYYLFZC022} has been empirically successful at learning high-level features that are robust to low-level noises (\eg, image rotation, flip). 
 As illustrated in Figure \ref{fig:firstpage}, a straightforward way to combine contrastive learning and recommendation is to construct positive/negative behavior sequences via random augmentation/sampling. Then a contrastive learning loss function, such as InfoNCE \cite{Oord_Li_Vinyals_2018}, helps to distinguish the relative difference between <query, positive> and <query, negative>. However, such a straightforward solution might be sub-optimal for recommendation due to the following two problems: 1) \textit{random augmentations} (\eg, item drop/replacement/re-order) on the user behavior sequence treat all items equally and cannot shield off the learned representation from false-positive interactions within implicit feedbacks; and 2) existing contrastive learning objectives that treat all positive/negative samples as equally important prevent the framework from modeling the \textit{nuances between augmented samples} and thus making it inferior in learning discriminating representations.
\end{sloppypar}

To investigate the above challenges and effectively consolidate the merits of contrastive learning into recommendation, we propose the \textit{\textbf{C}}ontrast over \textit{\textbf{C}}ontrastive \textit{\textbf{L}}earning framework for recommendation, abbreviated as CCL4Rec. CCL4Rec follows the sequential recommendation schema to have the behavior sequence as input. In essence, compared to existing contrastive learning techniques that solely consider the contrast on <positive, negative> pairs, CCL4Rec further considers the contrasts on <positive, positive>, <negative, negative> pairs. To make the contrast even more distinguishing, we enhance all kinds of contrasts with adaptive pulling/pushing strengths that correlate with the hardness/quality of augmented samples. In this way, CCL4Rec can learn to capture the nuances between the augmented samples in a \textit{contrast over contrast} manner. To achieve such adaptive or controllable contrasts, the traditional framework that constructs positive/negative views by random augmentation/sampling can be less trackable and effective. Towards this end, we propose to determine the hardness/quality of each augmented sample based on the \textit{importance} of replaced behaviors and the \textit{relatedness} of negative substitutes. The hardness scores permit not only adaptive pulling/pushing strengths but also controllable contrastive learning by deliberately choosing augmented samples with the desired hardness scores for contrasting. Therefore, the CCL4Rec framework is capable of modeling the nuances between different items within the historical behavior sequence (for the first problem), and the nuances between different augmented samples (for the second problem).

In the experiments, we equip CCL4Rec with a naive sequence modeling module and validate CCL4Rec on two large-scale micro-video recommendation benchmarks. Quantitative experiments, including ablation studies, and in-depth architecture analysis, demonstrate the strengths of CCL4Rec in being simple, lightweight, and effective compared with existing state-of-the-art methods. Remarkably, we show that CCL4Rec could achieve several orders of magnitude improvement on the training/inference speed with comparable performance to the state-of-the-art methods.
In a nutshell, the contributions of this work are listed as follows:
\begin{itemize}[leftmargin=*]
	\item We propose to learn discriminating user representations for micro-video recommendation via contrastive learning and further model the nuances of augmented samples in a contrast over contrast manner.
	\item We devise the novel CCL4Rec framework, which permits hardness-aware augmentation with trackable hardness scores for augmented samples, adaptive pulling/pushing strengths for different contrasting, and controllable contrastive learning that can choose augmented samples with the desired hardness scores for contrasting.
	\item We conduct extensive experiments which demonstrate the strengths of CCL4Rec in being simple, lightweight, and effective in learning discriminating user representations.
\end{itemize}

\section{Related Works}

\subsection{Contrastive Learning}

Recently, the contrastive learning objective has become advantageous for unsupervised representation learning and led to state-of-the-art results in various domains. The essence of contrastive learning lies in the infomax \cite{Linsker_1988} principle, where mutual information of a query and its positive views should be maximized, and the strategy of obtaining positive pairs. Typical augmentation methods for computer vision \cite{Chen_Kornblith_Norouzi_Hinton_2020} include random cropping and flipping \cite{Oord_Li_Vinyals_2018}. Tian \etal, \cite{Tian_Krishnan_Isola_2020} also use different views of the same scene as positive samples. Similarly, in natural language processing, Logeswaran and Lee \cite{Logeswaran_Lee_2018} use the context sentences as the positive views of the query sentence. Laskin \etal, \cite{Laskin_Srinivas_Abbeel_2020} employ the contrastive objective to improve sample efficiency of reinforcement learning. However, most of the existing contrastive learning frameworks treat all positive/negative samples as equally important, and we argue this can hinder the learned representation from being discriminating. 
In this paper, we propose to model the nuances between positives/negatives using contrast over the traditional contrastive learning. Also, the proposed controllable augmentation strategy constructs negatives with the desired hardness/quality from the query sample, which empirically leads to robust training and better performance.

Due to its effectiveness, there are increasing research interests on the intersection of contrastive learning and recommendation~\cite{DBLP:conf/sigir/ZhangYZC021,DBLP:conf/www/ZhangYYLFZC022}. 
It can be challenging to design augmentation strategies for recommendation samples since we have neither easily accessible contexts like the language nor straightforward (while effective) tools such as cropping or rotating like the computer vision. 
Noteworthy, Zhou \etal, \cite{Zhou_Ma_Zhang_Zhou_Yang_2020} directly treat items clicked by one user as the positive samples for contrastive learning and demonstrate that such an objective can help reduce popularity bias. Xie \etal, \cite{Xie_Sun_Liu_Gao_Ding_Cui_2020} construct positive views for one behavior sequence by cropping, masking, and reordering part of the sequence, and treat the other randomly sampled behavior sequences as negative samples. Their framework follows a typical pretraining strategy which constitutes a pretraining stage and a fine-tuning stage. Zhou \cite{Zhou_Wang_Zhao_Zhu_Wang_Zhang_Wang_Wen_2020} introduces the InfoNCE objective \cite{Oord_Li_Vinyals_2018} into the self-supervised masked item/attribute/segment prediction for sequential recommendation. Liu \etal, \cite{Liu_Ma_Ouyang_Xiong_2021} propose a graph-based framework that encapsulates graph perturbation, which can be viewed as the higher-order version of the masking operation in \cite{Xie_Sun_Liu_Gao_Ding_Cui_2020} and a debiased contrastive learning objective with items as positive/negative samples, similar to \cite{Zhou_Ma_Zhang_Zhou_Yang_2020}. Different from these works, we propose a hardness-aware augmentation strategy and construct positives/negatives directly from the original behavior sequence. The proposed CCL4Rec framework models the nuances between different behaviors and augmented samples, and thus learns discriminating user representations in the embedding hypersphere.

\subsection{Micro-video Recommendation}

Micro-video recommendation is a nascent research area in content-based recommender systems~\cite{DBLP:conf/mm/XunZZZZLH0C021} and attracts increasing research interests \cite{Liu_Chen_2019,Liu_Chen_Liu_Hu_2019,Chen_Liu_Zha_Zhou_Xiong_Li_2018} recently due to the booming of micro-video production and communication on the internet~\cite{DBLP:conf/mm/ZhangTYZKLZYW20,DBLP:conf/kdd/ZhangTZYKJZYW20}. Typically, Li \etal, \cite{Li_Liu_Yin_Cui_Xu_Nie_2019} propose to model the dynamic and multi-level interests using the temporal graph-based LSTM and a multi-level interest modeling layer. Wei \etal, \cite{Wei_Wang_Nie_He_Hong_Chua_2019} explicitly model the modality-specific user preferences and use graph neural networks to leverage inter-dependence between users and micro-videos in multiple modality-specific graphs. Hao \etal, \cite{Jiang_Wang_Wei_Gao_Wang_Nie_2020} devise the time-aware parallel masks for leveraging multi-scale time effects and the group routing algorithm to assign historical behaviors to different groups. Different from these works, we propose to learn discriminating user representations using contrastive learning, which remains largely unexplored in the literature of both generic and micro-video recommendations.

\section{Methods}

\subsection{Overview}

\begin{figure*}[t] \begin{center}
    \includegraphics[width=\textwidth]{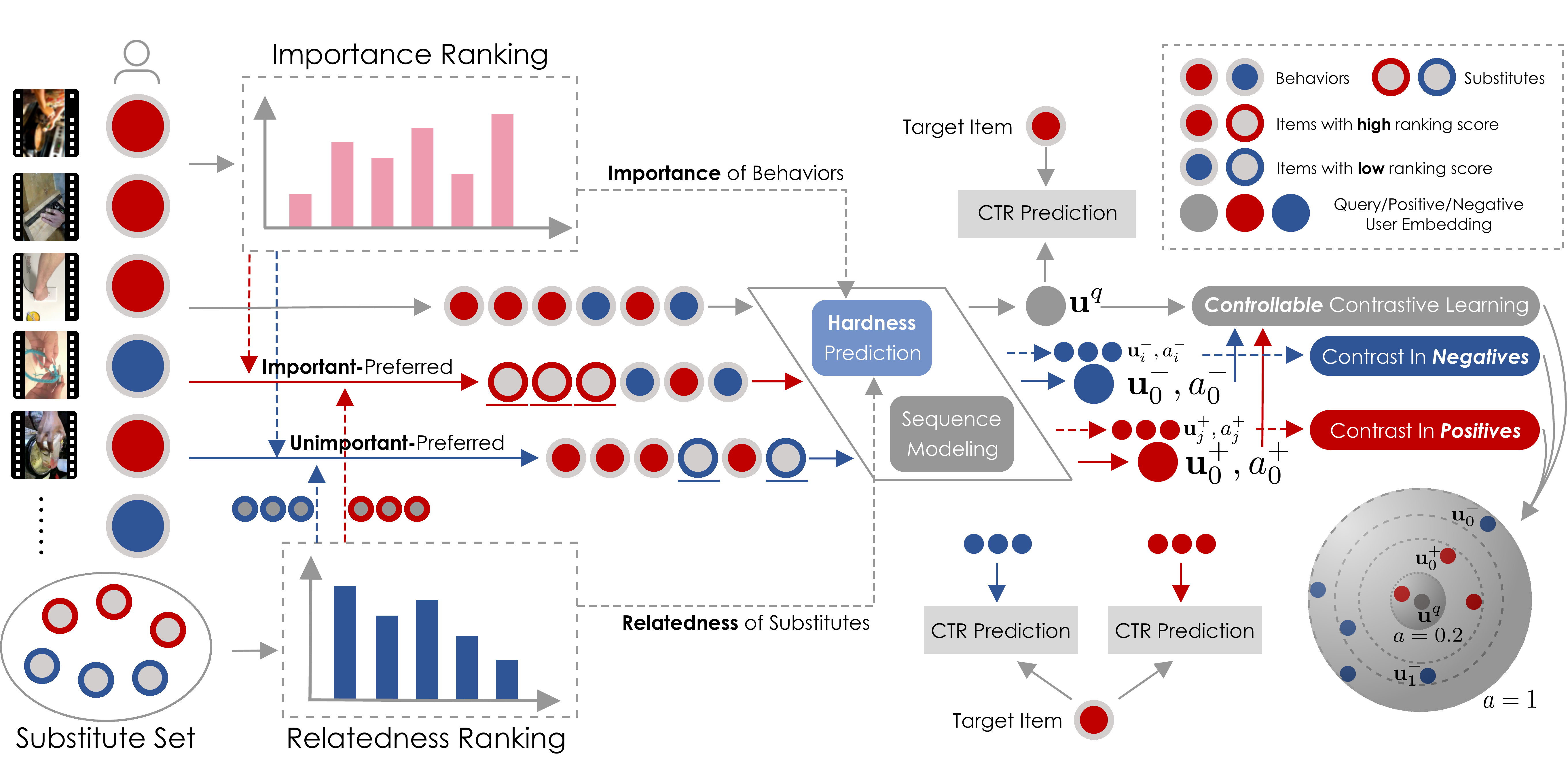}
    \caption{
    	Schematic of the proposed CCL4Rec architecture, which mainly encapsulates three critical contributions: 1) \textit{Hardness-aware augmentation}, that explicitly tracks the importance of behaviors and relatedness of substitutes in user behavior sequence augmentation, resulting in a hardness score of the augmented sample. 2) \textit{Contrast over contrast} objective that explicitly models the nuances between positives/negatives based on the hardness scores. 3) \textit{Controllable contrastive learning} that can deliberately choose augmented samples with the desired hardness scores.
	}
\label{fig:schema}
\end{center} \end{figure*}

We now present the contrast over contrastive learning framework for recommendation, termed CCL4Rec. The critical ideas of CCL4Rec are illustrated in Figure \ref{fig:schema}. CCL4Rec follows a schema of sequential recommendation, \ie, explicitly modeling the historical behavior sequence for prediction. 
Such a schema has the advantage of dynamically leveraging newly interacted items for prediction without model re-training. 
CCL4Rec mainly encapsulates three components: 1) \textit{Hardness-aware Augmentation}, which determines the importance of each historical behavior and the relatedness of each substitute. By jointly considering the importance and relatedness, we conduct replacement augmentation and obtain positive/negative user behavior sequences of various hardness scores. Controllable Augmentation means we can obtain augmented samples with the desired hardness scores.
Based on the scores, we can explicitly choose augmented samples of specific range of hardnesses along with the contrastive learning processes, \ie, controllable contrastive learning. Such scores further permit contrast over contrastive learning.
 2) \textit{Sequence Modeling}, which encodes the original behavior sequence and the augmented sequences into vectorial representations. We adopt a simplified design for efficiency. 3) \textit{Contrast over Contrast}, which explicitly models the nuances of different augmented samples. We achieve this by designing hardness-related objectives and contrasting either negative or positive samples besides the conventional <query, positive, negative> contrasting schema. Furthermore, based on the hardness of augmented samples, we learn from simple contrasts at the early stage of training and progressively learn from harder contrasts, which empirically leads to a more robust training process and better performance.

\subsection{Behavior/Substitute Ranking} Let $v$ and $u$ denote a micro-video and an user, and let the bold symbols $\mathbf{v}$ and $\mathbf{u}$ denote the corresponding vectorial representations. We denote the behavior sequence of $u$ as $X_u = \{ v_{u,t} \}_{t=1}^{N_u}$, where $v_{u,t}$ denotes the $t$th micro-video interacted by user $u$ and $N_u$ is the number of behaviors in the sequence. Since the following description is within the range of one user, we write $v_{t}$ in place of $v_{u,t}$ for brevity. Given $X_u$, we propose to determine the importance of each element $v_{t}$ in representing the interests of user $u$, \ie, behavior importance ranking. We firstly compute the relevance of each element with other items in the behavior sequence, \ie,
\begin{align}
	\alpha_{ij} = f_a(v_i, v_j) = (\mathbf{v}_i \mathbf{W}_1)^T (\mathbf{v}_j \mathbf{W}_2),
\end{align}
where $\mathbf{W}_1$ and $\mathbf{W}_2$ are trainable linear mapping matrices. We propose to sum the relevance scores to all items and treat the summed score as the importance for representing the user, \ie,
\begin{align}
	\alpha_{t} = f_s(v_t) = \sum_{j=1}^{N_u} \alpha_{tj},
\end{align}
Intuitively, when a micro-video is more relevant to other micro-videos watched by the user, it has a higher chance of being an important one. The computation of importance introduces trainable parameters, \ie, $\mathbf{W}_1$ and $\mathbf{W}_2$. Since we have no groundtruth for the importance score, we encapsulate these parameters into the sequence modeling module and jointly optimize them by the final cross-entropy loss function. We note that the proposed contrastive objectives that explicitly rely on the importance scores will not optimize $\mathbf{W}_1, \mathbf{W}_2$,
 and we achieve this simply using the \texttt{detach()} function in pytorch when computing the importance scores. We will illustrate the objectives and training in Section \ref{sec:obj} in detail.

For user $u$, all other items not in the behavior sequence will be treated as potential substitutes for replacement augmentation. To accommodate the hardness-aware augmentation described in Section \ref{sec:aug}, we propose to rank these substitutes and compute the relatedness scores, \ie, substitute relatedness ranking. However, it can be infeasible to rank the whole item gallery, which can easily reach a billion-scale in real-world recommender systems. Towards this end, we propose to approximate the above process and choose to rank a subset of randomly sampled substitutes, \ie, $Z_u = \{ v^z_k \}_{k=1}^{N_z}$, where $v^z_k$ is a potential substitute and $N_z$ is the number of sampled substitutes to be ranked. For user behaviors, we compute the importance score at the user level, and for substitutes, we compute the relatedness score at a fine-grained behavior level:
\begin{align}
	\beta_{tk} = f_b(v_t, v_k^z) = (\mathbf{v}_t \mathbf{W}_1)^T (\mathbf{v}_k^z \mathbf{W}_2),
\end{align}
where $\beta_{tk}$ denotes the relatedness of substitute $v^z_k$ and behavior $v_t$. We reuse the scoring matrices $\mathbf{W}_1$ and $\mathbf{W}_2$ to reduce parameters.

\subsection{Hardness-aware Augmentation} \label{sec:aug}

Based on the behavior importance scores and substitute relatedness scores, we propose to determine the \textit{hardness} of augmented sequences. The term \textit{\textbf{hardness}} is borrowed from the term \textit{hard negative mining} and refers to the quality of the mined samples \wrt the query sample. We note that we largely refrain from mining hard negatives from all other samples by augmenting the current sample to have positives/negatives. Augmentation has the potential to be efficient, controllable, and of high quality by manipulating samples at a fine-grained component level. We conduct replacement augmentation, \ie, $X^z_u = (X_u \setminus \bar X_u) \cup \bar Z_u $ by replacing $N_r$ micro-videos $\bar X_u$ with substitutes $\bar Z_u$ since replacement is more trackable as compared to other alternatives such as sequence re-order. 

\vpara{Constructing Negatives.} When we prefer to replace items with high relative importance score $\frac{\exp \left( \alpha_i \right)}{\sum_{v_j \in X_u} \exp \left( \alpha_j \right)}$, the constructed behavior sequences are potential negatives. We compute the hardness score for a potential negative $X^{z,-}_u$ as follows:
\begin{align}
	a^- = \sum_{(v_m, v_n^z)}^{N_r} \frac{\exp \left( \alpha_m \right)}{\sum_{v_i \in X_u} \exp \left( \alpha_i \right)} * \frac{\exp \left( \beta_{mn} \right)}{\sum_{v_j^z \in Z_u} \exp \left( \beta_{mj} \right)}, \label{eq:neghard}
\end{align}
where $v_m \in \bar X_u$ is a replaced behavior and $v_n^z \in \bar Z_u$ is the corresponding substitute. $\alpha_m$ and $\beta_{m,n}$ denote the importance score of $v_m$ for user $u$ and the relatedness score of substitute $v_n^z$ for $v_m$. We use softmax to transform the absolute importance and relatedness into the relative ones within the range of the behavior sequence and the substitute subset, respectively. The intuition behind the multiplication of relative importance and relatedness scores for constructing \textbf{\textit{negatives}} is that replacing an important item with a highly related substitute will make the augmented negative harder. For example, when a user loves watching the cat micro-videos, replacing a cat micro-video in the behavior sequence with a dog micro-video  (of high related score) will make the augmented behavior sequence much harder than replacing it with a cooking micro-video  (of low related score). We essentially treat all items the user does not click as negatives, which is a common strategy in training deep candidate generation models for recommendation. Also, replacing unimportant behaviors will intuitively lead to easier augmented samples.

\vpara{Constructing Positives.} When we prefer to replace items with high relative unimportance score $\frac{\exp \left( -\alpha_i \right)}{\sum_{v_j \in X_u} \exp \left( -\alpha_j \right)}$, the constructed behavior sequences are potential positives. We compute the hardness score for $X^{z,+}_u$ as follows:
\begin{align}
	a^+ = \sum_{(v_m, v_n^z)}^{N_r} \frac{\exp \left( - \alpha_m \right)}{\sum_{v_i \in X_u} \exp \left( - \alpha_i \right)} * \frac{\exp \left( - \beta_{mn} \right)}{\sum_{v_j^z \in Z_u} \exp \left( - \beta_{mj} \right)}. \label{eq:poshard}
\end{align}
The difference between positives and negatives lies in that we use the negative importance/relatedness score before softmax, which will result in a relative unimportance/unrelatedness score after softmax. The intuition behind the multiplication of relative unimportance and unrelatedness scores for constructing \textbf{\textit{positives}} is that replacing an unimportant item with an unrelated substitute might make the augmented positive be of high quality. For example, when a user dislikes ads micro-videos, replacing an ads micro-video in the behavior sequence with a micro-video that is far from ads  (of low related score) will have a higher chance of making the positive be of high quality than replacing it with another ads micro-video (of high related score). Replacing important behaviors will intuitively make the constructed positives be of low quality.

Based on the above computation, we can devise controllable augmentation by deliberately choosing the behaviors to be replaced and the substitutes to construct positives/negatives samples with the desired hardness scores, \ie, hardness-aware augmentation. The controllable augmentation permits many enticing contrastive techniques compared with the vanilla contrastive learning, of which the details are discussed in \ref{sec:obj}.

\subsection{Sequence Modeling} \label{sec:seqmodel}

Sequence modeling lies in the core of most sequential recommendation architectures, which transform the historical behavior sequence into a holistic (or multiple) user interest representation. 
To better demonstrate the effectiveness of the proposed framework, we propose to simplify the sequence modeling and introduce just one linear transformation layer $\mathbf{W}_3$ with \textbf{\textit{NO}} further trainable parameters. This design largely improves the efficiency of both training and inference, which is a critical merit for industrial systems. Specifically, we obtain the original user representation as:
\begin{align}
	\mathbf{\hat v}_{t} &= \sum_{i=1}^N \left( \mathbf{v}_i \mathbf{W}_3 \right) * \frac{\exp \left( \alpha_{ti} \right)}{\sum_{v_j \in X_u} \exp \left( \alpha_{tj} \right)},
\end{align}
\begin{align}
	\mathbf{u}^q &= \sum_{t=1}^N \mathbf{\hat v}_{t} * \frac{\exp \left( \alpha_t \right)}{\sum_{v_i \in X_u} \exp \left( \alpha_i \right)}.
\end{align}
where $\mathbf{v}_i$ is updated by aggregating information from all other micro-video in the behavior sequence by re-using the correlation score $\alpha_{ti}$ as the weights after softmax. The final user representation $\mathbf{u}^q$ sums all micro-video features by re-using the importance score $\alpha_t$ as the weights after softmax. This architecture has the strengths of being simple: efficient for training/serving, lower chance of overfitting, and less space complexity. Augmented behavior sequences $\{X^{z,+}_i\}_{i=1}^{N_p}, \{X^{z,-}_j\}_{j=1}^{N_n}$ are also transformed into the corresponding user representations $\{\mathbf{u}^+_i\}_{i=1}^{N_p}, \{\mathbf{u}^-_j\}_{j=1}^{N_n}$.

\subsection{Training and Objectives} \label{sec:obj}

\subsubsection{Contrast on Users} Based on the augmented positive and negative user representations, we conduct contrastive learning to learn discriminating user representations. A straightforward way is to employ the margin triplet loss function \cite{Balntas_Riba_Ponsa_Mikolajczyk_2016}:
\begin{align}
	\sum_{i=1}^{N_p} \sum_{j=1}^{N_n} \max \left( d\left(\mathbf{u}^{q}, \mathbf{u}^+_i\right)-d\left(\mathbf{u}^{q}, \mathbf{u}^-_j\right)+\delta, 0 \right),
\end{align}
where $d(\cdot)$ denotes the distance measurement and $\delta$ denotes the margin, which indicate the maximum relative distance between <original, positive> and <original, negative>.

\vpara{Contrast over Contrastive Learning.} However, the above formulation basically treats all positives and negatives as equally important. Failing to model the nuances between them might hinder contrastive learning from learning discriminating user representations. Towards this end, we propose to: 1) model the hardness of positives and negatives in the modeling; and 2) further contrast the negatives or positives in a contrast over contrast manner. Mathematically, the hardness can be incorporated into the objective as follows:
\begin{align}
	\delta^* &= \max(\min( (a^+_i + a^-_j) * \delta^s , \delta^u), \delta^l), \\
	\mathcal{L}_{ccl} &= \sum_{i=1}^{N_p} \sum_{j=1}^{N_n} \max \left( d\left(\mathbf{u}^{q}, \mathbf{u}^+_i\right)-d\left(\mathbf{u}^{q}, \mathbf{u}^-_j\right) + \delta^*, 0 \right),
\end{align}
where $a^+_i, a^-_j$ denote the hardness scores of augmented samples $\mathbf{u}^+_i, \mathbf{u}^-_j$, computed in Equation \ref{eq:poshard} and Equation \ref{eq:neghard}, respectively. Intuitively, if a positive is of high quality, we should pull the corresponding user representation $\mathbf{u}^+_i$ closer to the original $\mathbf{u}^{q}$. Similarly, if a negative is of high hardness, we should push $\mathbf{u}^-_j$ farther from the original. Therefore, the maximum relative distance, \ie, the margin $\delta$ should be larger to permit further contrasting than the other pairs. We use the sum of harness score $a^+_i + a^-_j$ to control the margin and employ hyper-parameters $\delta^s, \delta^u, \delta^l$ to control the scale, upper bound, lower bound of the final margin. Besides the conventional way of contrasting positives with negatives, we further contrast positives or negatives as follows:
\begin{align}
	\delta^{+,*} &= \max(\min( (a^+_i - a^+_k) * \delta^s , \delta^u), \delta^l), \\
	\mathcal{L}_{ccl}^+ &= \sum_{i=1}^{N_p} \sum_{a^+_k < a^+_i} \max \left( d\left(\mathbf{u}^{q}, \mathbf{u}^+_i\right)-d\left(\mathbf{u}^{q}, \mathbf{u}^+_k\right) + \delta^{+,*}, 0 \right),
\end{align}
Unlike $\mathcal{L}_{ccl}$, we use $a^+_i - a^+_k$ to control the margin. Intuitively, when contrasting positives, if the qualities of two augmented samples largely differ from each other, the maximum relative distance should be large. Similarly in spirit, we could compute $\mathcal{L}_{ccl}^-$ by contrasting the negatives. We note that the above objectives that explicitly rely on the hardness scores will not optimize the score-related parameters.

\vpara{Controllable Contrastive Learning.} As demonstrated in Section \ref{sec:aug}, with the computation of hardness scores, we can easily have hardness-aware augmentation, which further permits controllable contrastive learning by deliberately choosing augmented samples of the desired hardness scores for contrasting. In this paper, we propose multiple controllable contrast strategies and discuss their empirical performance in Section \ref{sec:ccs}. We take the \textit{easy2hard} strategy as an example for illustration. With the easy2hard strategy, we propose to learn from augmented samples with smaller hardness scores at the beginning and increase the hardness scores of the chosen samples as the learning procedure processes. Specifically, when constructing positives/negatives, we filter the half items with the largest/smallest importance scores $\alpha_i$, and further sample half items from the filtered set to be replaced, \ie, $\bar X_u$. The sampling probabilities are initially set to the relative unimportance scores, \ie, $\frac{\exp \left( -\alpha_i \right)}{\sum_{v_j \in X_u} \exp \left( -\alpha_j \right)}$, which means we construct positives with lower hardness scores at the early stage of training. The sampling probabilities will be linearly transformed towards the importance scores $\frac{\exp \left( \alpha_i \right)}{\sum_{v_j \in X_u} \exp \left( \alpha_j \right)}$ along with the training processes. Similarly, we sample substitutes according to the unrelatedness scores $\frac{\exp \left( - \beta_{ik} \right)}{\sum_{v_j^z \in Z_u} \exp \left( - \beta_{ij} \right)}$, which will be linearly transformed towards the relatedness score along with the training. Controllable negative augmentation shares similar processes as the above.

\subsubsection{click-through-rate prediction.} As a common practice, we incorporate a multi-layer perceptron (MLP) to predict the CTR and a cross-entropy loss as the objective:
\begin{align}
	\hat{y} &= \operatorname{MLP}\left( \left[\mathbf{u}, \mathbf{v}_i\mathbf{W}_4, \mathbf{u} * \mathbf{v}_i\mathbf{W}_4 \right] \right), \\
	\mathcal{L}_{ce} &= -\sum_{\hat{y}, y^{*}} y^{*}\log \left(\hat{y} \right)+\left(1-y^{*}\right) \log \left(1-\hat{y}\right),
\end{align}
where $\left[ \cdot \right]$ denotes the concatenation operation and $*$ denotes element-wise product. We also compute the corresponding cross-entropy loss for augmented samples. While $\mathcal{L}_{ce}^+$ for positives share the same computation, $\mathcal{L}_{ce}^-$ is computed with the groundtruth $y_{u i}^{*}$ reverted, \ie, $1 \rightarrow 0$, and disregards target micro-videos with label 0.

\subsubsection{Contrast on User-Item Pairs} Besides contrasting users, we also consider the contrast between the user and positive/negative items. The objective function can be written as:
\begin{align}
	\mathcal{L}_{cui} = - \sum_u \log \frac{\sum_{\hat{y}_u, y_u^{*}} y_u^* \exp \left(  \hat{y}_u \right)}{ \sum_{\hat{y}_u, y^{*}_u} y^*_u \exp \left(  \hat{y}_u \right) + \sum_{\hat{y}_u, y^{*}_u} (1-y^*_u) \exp \left(  \hat{y}_u \right)}, 
	\label{eq:cont}
\end{align}
This contrast is an essential complement to the original cross-entropy loss by globally enlarging the gap between the summed predictions of positive pairs and that of negative pairs.

\subsubsection{Training}

During training, we naively sum the above loss functions and do not tune the relative weights with bells and whistles to better demonstrate the strengths of the proposed objectives:
\begin{align}
	\mathcal{L} = \mathcal{L}_{ccl} + \mathcal{L}_{ccl}^- + \mathcal{L}_{ccl}^+ + \mathcal{L}_{ce} + \mathcal{L}_{ce}^+ + \mathcal{L}_{ce}^- + \mathcal{L}_{cui}.
\end{align}

\section{Experiments}

We are concerned with the following research questions:

\begin{itemize}[leftmargin=*]
	\item \textbf{RQ1}: How does CCL4Rec perform compared with the state-of-the-art micro-video recommenders?
	\item \textbf{RQ2}: Do all contrastive objectives contribute to the effectiveness of CCL4Rec? How do different key hyper-parameters settings and controllable contrastive strategies affect the performance?
	\item \textbf{RQ3}: Does CCL4Rec learn discriminating user presentations?
\end{itemize}

\subsection{Experimental Settings}

\begin{table}[!t]
\caption{Statistics of the Datasets.}
\setlength{\tabcolsep}{3pt}
\centering
\begin{tabular}{l cc cc}
\toprule
Dataset & \#Users & \#Items & \#Interactions & \#Density  \\
    \midrule
    MicroVideo-1.7M		& 	10, 986  &   1, 704, 880  &  12, 737, 619  & 0.068\% 	 \\ 
    Kuaishou	& 	10, 000 & 	 3,239,534 & 13, 661, 383	 & 0.042\%	 \\
    \bottomrule
\end{tabular}%
    \label{tab:staData}
\end{table}
%

\vpara{Datasets} As a common practice of micro-video recommendation research \cite{Jiang_Wang_Wei_Gao_Wang_Nie_2020,Li_Liu_Yin_Cui_Xu_Nie_2019}, we conduct experiments on two micro-video datasets, \ie, MicroVideo-1.7M \cite{Chen_Liu_Zha_Zhou_Xiong_Li_2018} and Kuaishou \footnote{ \url{https://www.kesci.com/home/competition/5ad306e633a98340e004f8d1}.}. These two datasets both provide pre-extracted video features and logged interactions in the form of <User ID, Video ID, Timestamp>. Each interaction is labeled as positive or negative according to whether the user clicked the exposed video when he/she saw the thumbnail. We exactly follow the data pre-processing of Chen \etal, \cite{Chen_Liu_Zha_Zhou_Xiong_Li_2018}, and Li \etal, \cite{Li_Liu_Yin_Cui_Xu_Nie_2019} for a fair comparison with the existing state-of-the-art models. The statistics of these datasets are listed in Table \ref{tab:staData}.

\vpara{Evaluation Protocals} To conduct a comprehensive evaluation on CCL4Rec and multiple comparison methods, we employ multiple numerical metrics, \ie, Area Under Curve (AUC), Precision, Recall, and F1-score, which are widely used in micro-video recommendation \cite{Jiang_Wang_Wei_Gao_Wang_Nie_2020,Li_Liu_Yin_Cui_Xu_Nie_2019} and can reveal the effectiveness of models in multiple aspects. 
Specifically, Precision indicates the fraction of recommended items that are actually clicked by users among the number of recommended items in total while Recall cares those among the number of clicked items in total. F1 score is computed as the harmonic mean of Precision and Recall, and thus taking these two metrics into account. AUC indicates the probability of a randomly sampled item that is actually clicked being ranked higher than a randomly sampled item that is not.
 We report the results that are computed based on the Top 50 recommended items by all models, \ie, Precision@50, Recall@50, and F1-Score@50.

\begin{table*}[h]
\centering
    \caption{Overall performance comparison between CCL4Rec and micro-video recommendation baselines. We highlight the best/next-best performance with bold/underline style. We show that CCL4Rec achieves comparable performance to the state-of-the-art methods but is of far more efficiency as illustrated in Table \ref{tab:complex}.}
{\setlength{\tabcolsep}{0.9em}\renewcommand{\arraystretch}{1.1}\begin{tabular}{l cc cc cc cc}
\toprule
&\multicolumn{4}{c}{ MicroVideo-1.7M } & \multicolumn{4}{c}{ Kuaishou } \\
\cmidrule(lr){2-5}\cmidrule(lr){6-9}
Model & AUC & Precision@50 & Recall@50 & F1-Score@50  & AUC & Precision@50 & Recall@50 & F1-Score@50 \\
    \midrule
    BPR       & 0.583 & 0.241 & 0.181 & 0.206 & 0.595 & 0.290 & 0.387 & 0.331	 		\\ 
    LSTM       & 0.641 & 0.277 & 0.205 & 0.236 & 0.713 & 0.316 & 0.420 & 0.360 		\\ 
    CNN      & 0.650 & 0.287 & 0.214 & 0.245 & 0.719 & 0.312 & 0.413 & 0.356		\\ 
    NCF      & 0.672 & 0.316 & 0.225 & 0.262 & 0.724 & 0.320 & 0.420 & 0.364		\\ 
    ATRank 	       & 0.660 & 0.297 & 0.221 & 0.253 & 0.722 & 0.322 & 0.426 & 0.367		\\  
    THACIL         & 0.684 & \textbf{0.324} & 0.234 & 0.269 & 0.727 & 0.325 & 0.429 & 0.369		\\ 
    ALPINE        & 0.713 & 0.300 & 0.460 & 0.362 & 0.739 & 0.331 & 0.436 & 0.376	\\  
    MTIN          &	 \textbf{0.729}	 &	\underline{0.317}	 &	\textbf{0.476}	 &	\textbf{0.381}	  &		\textbf{0.752}	 &	 \textbf{0.341}	 &	\textbf{0.449}	 &	 	\textbf{0.388}	\\ 
    \midrule
    CCL4Rec            &	\underline{0.722}	 &	0.312 &	0.472 &	0.376	 &	\underline{0.750} &	\underline{0.340}	&	\textbf{0.449}	 &	\underline{0.387}		\\   
    \bottomrule
\end{tabular}}
    \label{tab:comparison}
\end{table*}

\begin{table}[!t]
\centering
    \caption{Computation Complexity Analysis. Training time concerns one epoch and testing concerns one batch. We choose two best-performing state-of-the-art methods and all models are with batch size of 32 and 1 NVIDIA V100 GPU. We use the official codebases of ALPINE and MTIN provided by the authors and run them with default settings on the Kuaishou dataset. We highlight the speedup rate over two state-of-the-art methods.}
{\setlength{\tabcolsep}{0.7em}
\begin{tabular}{l ccc}
\toprule

   & Training (s) & Inference (s) & \#Parameters   \\
    \midrule
    ALPINE    &  	10,778.50   &  	 0.7191  & 	    208,854,467    \\
    MTIN      &  	2,896.83    &  	3.2070    &   10,849,706	       \\
    \midrule
    CCL4Rec   &  	121.88    &  0.0076	    &   	  758,341    \\
   \textbf{Efficiency}   &  		\textbf{88.44}/\textbf{23.8}$\times$    &  	 \textbf{94.6}/\textbf{463.4}$\times$   &   	  \textbf{275.4/14.3}$\times$    \\
    \bottomrule
\end{tabular}}%

    \label{tab:complex}
\end{table}

\vpara{Implementation Details} We use Adam optimizer \cite{Kingma_Ba_2015} for training CCL4Rec. All experiments are with batch size 32 and learning rate 0.003. We use weight normalization at a rate of $1e-7$. We randomly sample 10,000 micro-videos from the whole gallery as potential substitutes shared by the users in one batch. \ie. $N_z=10,000$. We disregard the trainable user embedding layer in our experiment. Micro-video embeddings are 128-dimensional vectors. We tune $\delta^l, \delta^u$ among $\{0.5, 0.6, 0.7, 0.8, 0.9\}$ and $\{1.1, 1.2, 1.3, 1.4, 1.5\}$, respectively, and empirically choose the ones that achieve best results. We choose $\delta^u$ that can roughly map most scores into the range of $[\delta^l, \delta^u]$. For a fair comparison with the state-of-the-art micro-video recommender, \ie, MTIN \cite{Jiang_Wang_Wei_Gao_Wang_Nie_2020}, which uses pretrained user interest embeddings in their experiment, we also incorporate the embeddings provided by the authors for better performance. Specifically, there are six user interest embeddings for each user, and we model them using the sequence modeling module described in Section \ref{sec:seqmodel} with separate linear transformation matrices. We also model the micro-videos that are recommended to users but not watched by them as an additional sequence using the proposed sequence modeling module, following ALPINE \cite{Li_Liu_Yin_Cui_Xu_Nie_2019} and MTIN \cite{Jiang_Wang_Wei_Gao_Wang_Nie_2020}. We note that the proposed augmentations are not performed on these additional sequences, and we obtain the final CTR prediction by the average of the predictions from multiple sequences.

\subsection{Comparison Methods}

To demonstrate the effectiveness of CCL4Rec, we consider the following state-of-the-art recommenders as comparison methods.

\begin{itemize}[leftmargin=*]
	\item \textbf{BPR} \cite{Rendle_Freudenthaler_Gantner_Schmidt_Thieme_2009}. Bayesian personalized ranking devises the maximum posterior estimator from the Bayesian perspective. The BPR loss is computed as the relative difference between positive pairs and negative pairs.

	\item \textbf{LSTM} \cite{Zhang_Dai_Xu_Feng_Wang_Bian_Wang_Liu_2014}. An LSTM based sequential recommender. We use the final hidden state as the user interest vector and an MLP predictor for click-through-rate prediction.

	\item \textbf{CNN}. CNN extracts features from the behavior sequence, and we take the pooled feature as the user interest vector.
We use various kernels with various window sizes to extract features from the behavior sequence and apply global pooling on the extracted features to obtain user interest vector. Similarly, we use an MLP predictor for prediction.

	\item \textbf{NCF} \cite{He_Liao_Zhang_Nie_Hu_Chua_2017}. NCF is a collaborative filtering recommender equipped with deep neural networks, which can be of great representation power compared with the traditional inner product.

	\item \textbf{ATRank} \cite{Zhou_Bai_Song_Liu_Zhao_Chen_Gao_2018}. ATRank is a ranking recommender that comprehensively leverages the power of attention mechanisms.

	\item \textbf{THACIL} \cite{Chen_Liu_Zha_Zhou_Xiong_Li_2018}. THACIL divides the historical behaviors into multiple temporal blocks and captures both the intra-block and inter-block correlations. In addition, THACIL considers both category-level and item-level user interests for representing users' interests.

	\item \textbf{ALPINE} \cite{Li_Liu_Yin_Cui_Xu_Nie_2019}. The essence of ALPINE lies in the temporal graph-based LSTM module, which captures dynamic interests within the temporal behavior graph, and the multi-level interest modeling layer, which models multi-type behaviors.

	\item \textbf{MTIN} \cite{Jiang_Wang_Wei_Gao_Wang_Nie_2020}. To investigate the multi-scale time effects and item-to-interest grouping problem, MTIN devises the time-aware parallel masks and the group routing algorithm.
\end{itemize}

\begin{sloppypar}
We report the evaluation results of all methods on two micro-video recommendation benchmarks in Table \ref{tab:comparison}. Since computation complexity is one of the critical factors considered by industrial recommender systems, we also report the training/inference/\#parameters of the strongest state-of-the-art methods (\ie, ALPINE, MTIN) and CCL4Rec in Table \ref{tab:complex}. By analyzing these results, we have the following observations:
	\begin{itemize}[leftmargin=*]
	\item Although CCLRec cannot beat the strongest state-of-the-art method, \ie, MTIN, in most cases, it achieves comparably good results and outperforms many other existing micro-video recommendation state-of-the-art methods. In particular, the relative improvement over other simple designs (\eg, BPR, LSTM, CNN, NCF) and also some advanced approaches (\eg, ATRank, THACIL) reach nearly 100\% and 8\% with Recall@50 on the MicroVideo and Kuaishou benchmarks, respectively. We attribute these advantages to the following aspects: 1) The contrastive learning objective helps the sequence modeling module to learn more distinguishing user representations that can mitigate the effects of ubiquitous false-positive interactions by contrasting the learned representation with augmented samples. 2) The proposed hardness-aware augmentation considers the importance of replaced items and the relatedness of substitutes, and accordingly determining the hardness scores of augmented samples in a fine-grained manner. These designs help to model the nuances of different items in augmentation, and the nuances of different augmented samples in contrasting, leading to even more distinguishing user representations.	
	\item While achieving competitive performance, CCL4Rec significantly improves the training efficiency (88.44/23.8 times faster than ALPINE/MTIN) and the inference efficiency (94.6/463.4 times faster than ALPINE/MTIN). These results demonstrate the strengths of CCL4Rec in being simple and effective, which are critical for industrial scenarios where the number of users/items can easily reach billions in scale. Furthermore, the number of parameters is orders of magnitude (275.4/14.3) less than the other two state-of-the-art methods. Considering the rather simplified sequence modeling module in the CCL4Rec framework, CCL4Rec has the potentials to enhance a tiny-size model that will be used for on-device/edge computing. 
\end{itemize}
\end{sloppypar}

\subsection{Study of CCL4Rec (RQ2)}

\begin{table}[!t]
\centering
    \caption{Analysis on the number of augmented positives/negatives.}
{\setlength{\tabcolsep}{0.45em}
\begin{tabular}{l cccc}
\toprule
$N_p, N_n$
   & AUC & Precision@50 & Recall@50 & F1-Score@50   \\
    \midrule
    2    & 0.7425 & 0.3386 & 0.4469 & 0.3852 	       \\
    3    & 0.7471 & \textbf{0.3402} & \textbf{0.4491} & \textbf{0.3871} 	      \\
    4    & \textbf{0.7503} & 0.3401 & 0.4487 & 0.3869 	     \\
    5    & 0.7487 & 0.3397 & 0.4479 & 0.3863 	     \\
    \bottomrule
\end{tabular}}%

    \label{tab:numposneg}
\end{table}

\begin{table}[!t]
\centering
    \caption{Analysis on different contrastive learning objectives.}
{\setlength{\tabcolsep}{0.45em}
\begin{tabular}{l cccc}
\toprule
Model
   & AUC & Precision@50 & Recall@50 & F1-Score@50   \\
    \midrule
    CCL4Rec    &  	\textbf{0.7503}   &  	\textbf{0.3401}   &  	\textbf{0.4487}  & \textbf{0.3869}	       \\
    - $\mathcal{L}_{ccl}^{-/+}$    &  0.7388	    &  0.3359	    &   0.4439	   &  	0.3824     \\
    - $\mathcal{L}_{ccl}$    &  0.7273	    &  	 0.3353   &   	0.4428   &  0.3816	      \\
    - $\mathcal{L}_{cui}$    &  	 0.7239   & 	 0.3327    &   0.4404	    &  	 0.3790   \\
    \bottomrule
\end{tabular}}%

    \label{tab:contrastObj}
\end{table}

\subsubsection{Study of the controllable contrastive strategies.} \label{sec:ccs} 
\begin{figure}[!t] \begin{center}
\begin{subfigure}{.45\textwidth}
	\includegraphics[width=1\linewidth]{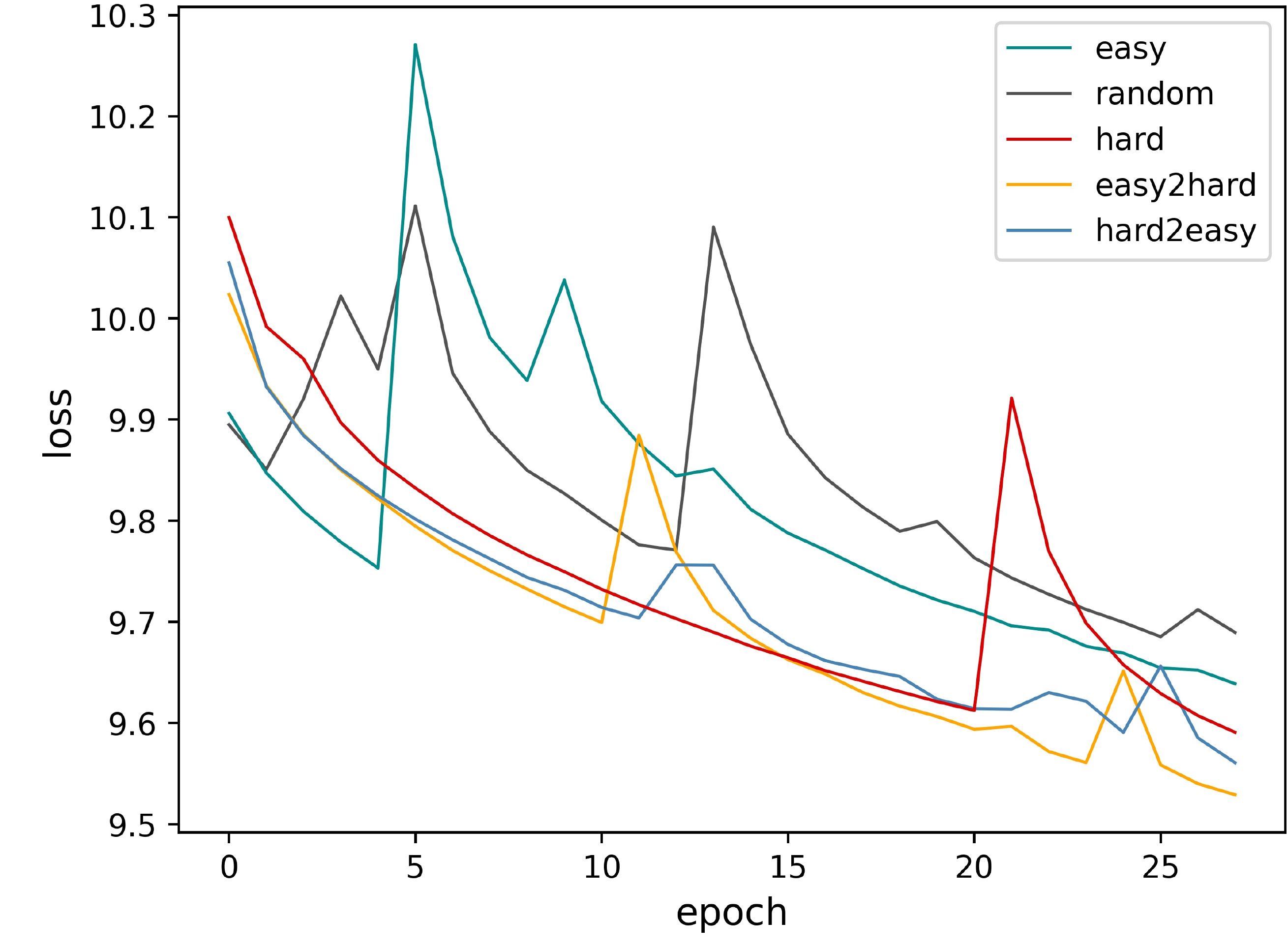}
    \caption{
    \footnotesize{Training loss}
    	}
\label{fig:tsnebase}
\end{subfigure}
\begin{subfigure}{.45\textwidth}
	\includegraphics[width=1\linewidth]{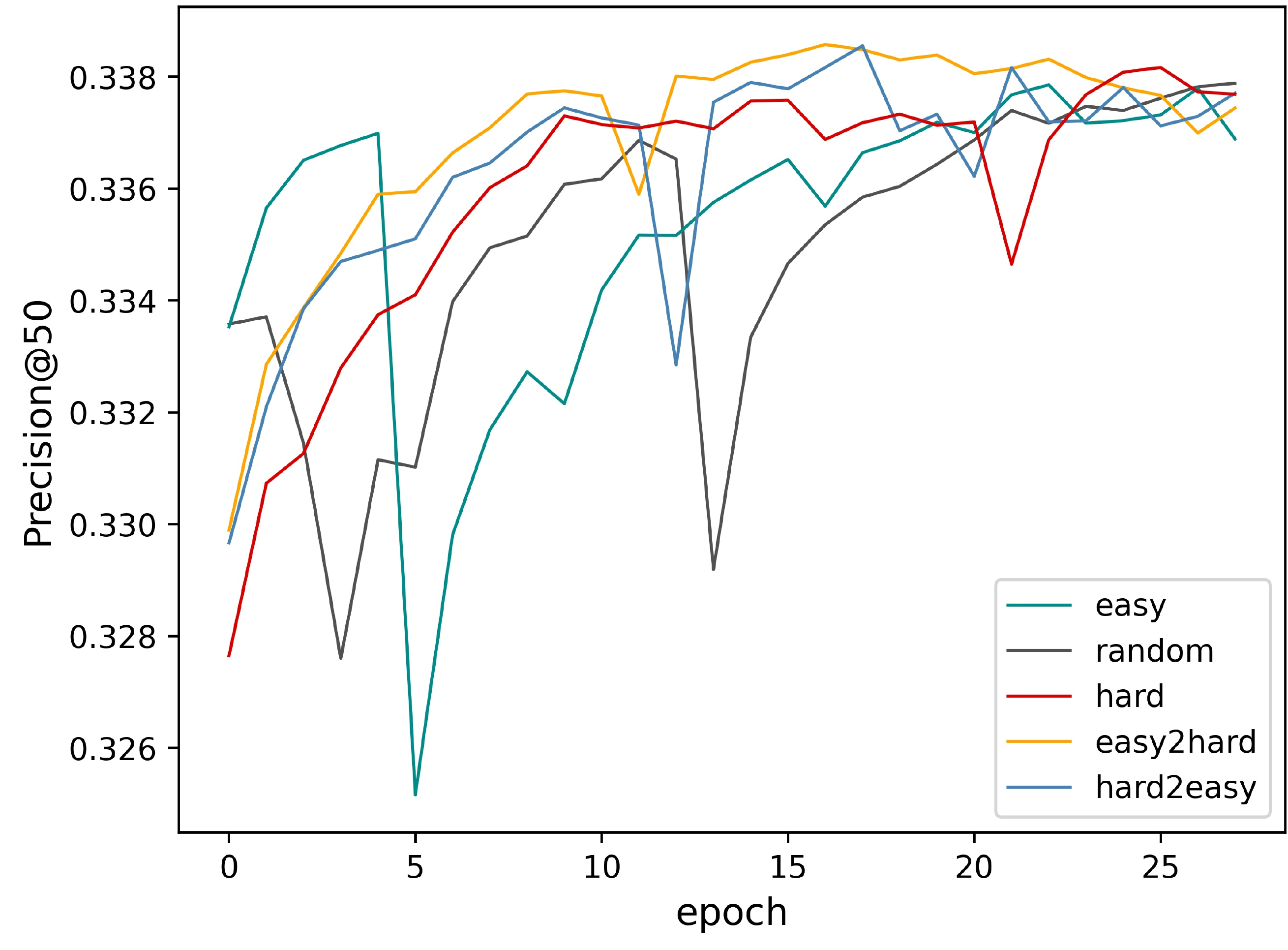}
    \caption{
   \footnotesize{Testing Precision}
    	}
\label{fig:tsneDemure}
\end{subfigure}
    \caption{
    Training curves (training loss and testing precision) of different controllable contrastive strategies.
    	}
    \label{fig:control}
\end{center} \end{figure}

To demonstrate the merits of controllable contrastive learning empowered by the controllable augmentation, we devise multiple strategies as follows: 
\begin{itemize}[leftmargin=*]
	\item The \textit{harder} strategy, which means we prefer to choose the augmented samples with larger hardness scores. We achieve this by using importance/relatedness scores as weights to sample the micro-videos to be replaced and the substitutes;
	\item The \textit{easier} strategy, which means we prefer ones with smaller hardness scores. We achieve this by using unimportance / unrelatedness scores as weights.
	\item The \textit{easy2hard} strategy as illustrated in Section \ref{sec:obj};
	\item The \textit{hard2easy} strategy; and 5) The \textit{random} strategy as a widely used baseline.
\end{itemize}
From the training curve depicted in Figure \ref{fig:control}, we have the following observations: 1) compared to the baseline \textit{random} strategy that has several high loss/performance jitters, other controllable contrastive strategies overall have significantly less and lower jitters; 2) among all the contrastive strategies, the \textit{easy2hard} strategy has the most robust training process and yields the best performance in most times. We note that $random$ strategy can be the most straightforward one that is widely used in many contrastive learning based methods. However, according to our analysis, it might lead to unstable training process and more training epochs to converge. These results jointly demonstrate the necessity of controllable contrastive learning and the effectiveness of our design in boosting performance and training robustness.

\subsubsection{Study of the number of augmented positives/negatives.} We vary the number of augmented positives/negatives to reveal its effect on the performance. The results in Table \ref{tab:numposneg} indicate that: 1) the performance of CCL4Rec is overall insensitive to this hyper-parameter and small $N_p, N_n$ will lead to competitive performance; 2) increasing $N_p, N_n$ ($2 \rightarrow 3,4$) will yield performance improvement since the contrast over contrast objective should learn more distinguishing user representation by modeling the nuances between multiple positives/negatives; 3) Further increasing $N_p, N_n$ ($3 \rightarrow 4$) will lead to fewer gains. The reason for this might be that moderate $N_p, N_n$ is enough to reach the optimal, which can be computationally efficient.

\subsubsection{Study of the contrastive objectives. (Ablation Study)} The critical contributions of CCL4Rec are the contrastive objectives. To evaluate whether each of them contributes to the final performance, we ablate CCL4Rec by progressively removing them and test the resulting architectures. Specifically, we observe the following from Table \ref{tab:contrastObj}: 1) By removing $\mathcal{L}_{ccl}^{-/+}$, which means we disregard the contrasts between positives or negatives, the performance gap between the resulting model and the full model demonstrates the necessity and effectiveness of the contrast over contrast framework; 2) By further removing $\mathcal{L}_{ccl}$, \ie, eliminating the effect of contrasting the learned representation with positive and negative ones, we can observe a clear performance drop. It is noteworthy that $\mathcal{L}_{ccl}$ also exhibits a contrast over contrast merit since the adaptive pulling/pushing strengths help CCL4Rec model the nuances between different augmented samples; 3) further removing $\mathcal{L}_{cui}$, which means the CCL4Rec framework totally lost the contrastive learning capability. The performance drop further shows the merits of the proposed contrastive learning framework.

\subsection{Case Study (RQ3)}

\begin{figure}[!t] \begin{center}
\begin{subfigure}{.245\textwidth}
	\includegraphics[width=1\linewidth]{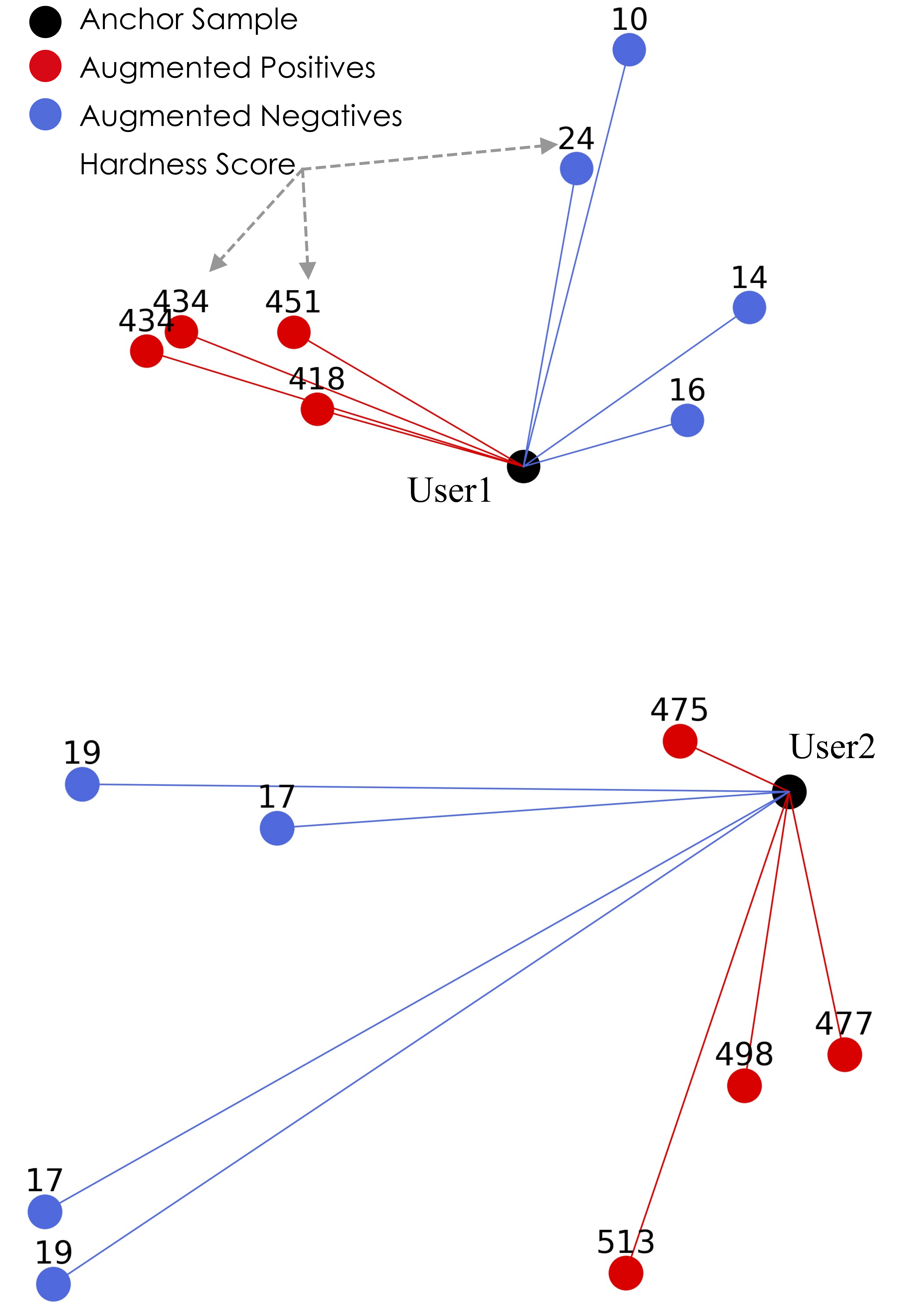}
    \caption{
    \footnotesize{Contrastive baseline}
    	}
\label{fig:tsnebase}
\end{subfigure}
\begin{subfigure}{.227\textwidth}
	\includegraphics[width=1\linewidth]{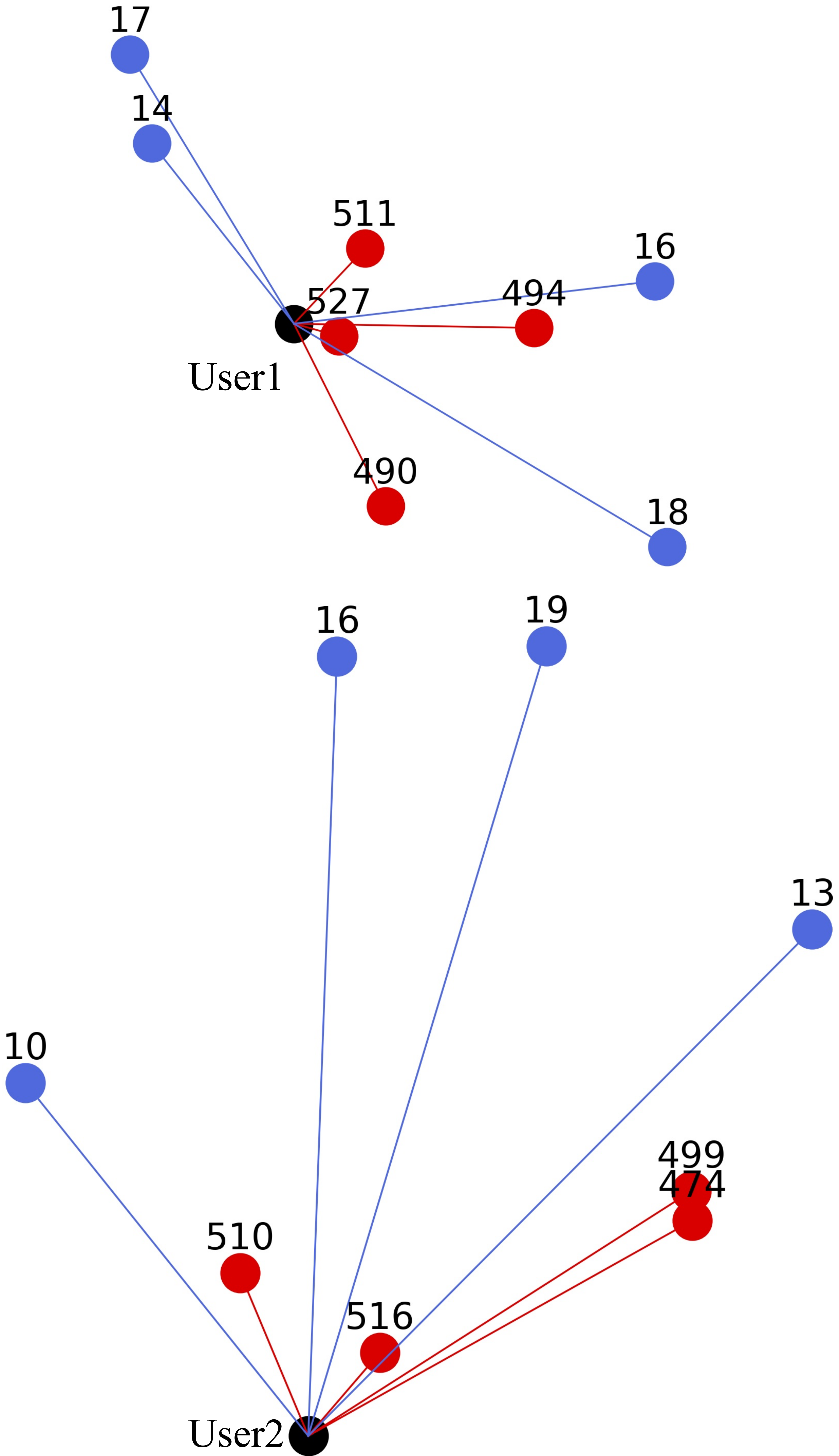}
    \caption{
   \footnotesize{Contrast over Contrast}
    	}
\label{fig:tsneDemure}
\end{subfigure}
    \caption{
    Case study with learned t-SNE transformed representations derived from the contrastive baseline and our CCL4Rec framework. Black/red/blue nodes represent the query/positive/negative samples. Each augmented sample is with a hardness score predicted by the corresponding model.
    	}
    \label{fig:case}
\end{center} \end{figure}

To evaluate whether we learn discriminating user representations via contrast over contrastive learning, we follow a widely used case study schema \cite{Wang_He_Wang_Feng_Chua_2019,Wei_Wang_Nie_He_Chua_2020} to visualize the t-SNE transformed embeddings. We adopt a contrastive baseline that is trained with $\mathcal{L}_{ce}$ and $\mathcal{L}_{ccl}$ without adaptive pulling/pushing strengths, \ie, the architecture without contrast over contrast. We plot two randomly sampled users and each user has four augmented positives/negatives with the hardness scores predicted by the corresponding model. As shown in Figure \ref{fig:case}, we have the following observations: 
\begin{itemize}[leftmargin=*]
	\item Both results yield clear gap between positives and negatives, which reveal the effectiveness of contrastive learning.
	\item The positives/negatives are less distinguishable from each in the baseline hyper-embedding space. Since positives and negatives are obtained by transforming important/unimportant items in the behavior sequence, these results basically indicate that the contrastive baseline treat all items as equally important and thus learning indiscriminating representations.
	\item By further considering the hardness scores, CCL4Rec successfully pushes negatives with high hardness scores farther and pulls positives with high hardness scores closer (while the baseline can not). These results demonstrate the effectiveness of contrast over contrast objective in CCL4Rec.
\end{itemize}

\balance

\section{Conclusion and Future Work} \label{sec:conclusion}

In this work, we investigate the problem of learning discriminating user representation via contrastive learning. Different from existing contrastive learning approaches, we propose to better accommodate recommendation using hardness-aware augmentation and contrast over contrastive learning. By hardness-aware augmentation, we can easily obtain augmented samples with the desired hardness scores by deliberately manipulating the importance of behaviors being replaced and the relevance of substitutes. Based on the hardness scores, we devise the controllable contrastive strategies, contrasts on negatives or positives, and contrast with adaptively pulling/pushing strengths, which jointly help to model the nuances between different augmented samples and thus learning discriminating representations.

To the best of our knowledge, the contrast over contrastive learning framework is among the earlest attempts in research on recommendation, while also contributing to the literature of generic contrastive learning. We believe the insights of CCL4Rec are inspirational to future developments on learning effective user representations and may be beneficial to a broad range of research domains. We plan to further explore the strengths by designing better augmentation cost prediction modules and better objectives to model the nuances of augmented samples. Another future direction is to take a more in-depth analysis of controllable contrastive learning. We mainly analyze the effects on the training process and final performance. Carefully designed strategies and broader merits are ripe for exploration. Some previous works found that the contrastive learning objective can be beneficial for bias reduction. We thus plan to also explore whether the contrast over contrastive learning framework can further help to reduce biases in recommender systems (\eg, popularity bias) compared with the traditional one.

\newpage
\newpage
\balance
\bibliographystyle{ACM-Reference-Format}
\bibliography{sections/9.citations}

\end{document}